\documentclass[twocolumn,floatfix,prb,showpacs,superscriptaddress]{revtex4}

\usepackage{color}
\usepackage{array}
\usepackage{amsmath}%
\usepackage{amssymb}
\usepackage{wasysym}
\usepackage{bm}
\usepackage{graphicx}
\usepackage{txfonts}
\usepackage{epsfig,hyperref}

\newcommand{\ket}[1]{|\, #1 \, \rangle}

\newcommand{\dg}{\dag}

\newcommand{\be}{\begin{equation}}
\newcommand{\ee}{\end{equation}}
\newcommand{\bc}{\begin{center}}
\newcommand{\ec}{\end{center}}

\begin{document}
\title{Engineering Quantum Hall Phases in Synthetic Bilayer Graphene system}
\author{Ze-Pei Cian}
\affiliation{Department of Physics,University of Maryland, College Park, Maryland 20742, USA}
\affiliation{Joint Quantum Institute, NIST/University of Maryland, College Park, Maryland 20742, USA}
\author{Tobias Grass}
\affiliation{Department of Physics,University of Maryland, College Park, Maryland 20742, USA}
\affiliation{Joint Quantum Institute, NIST/University of Maryland, College Park, Maryland 20742, USA}
\affiliation{ICFO-Institut de Ciencies Fotoniques, The Barcelona Institute of Science and Technology, 08860 Castelldefels (Barcelona), Spain}
\author{Abolhassan Vaezi}
\affiliation{Department of Physics, Sharif University of Technology, Tehran 14588-89694, Iran}
\author{Zhao Liu}
\affiliation{Zhejiang Institute of Modern Physics, Zhejiang University, Hangzhou 310027, China}
\author{Mohammad Hafezi}
\affiliation{Department of Physics,University of Maryland, College Park, Maryland 20742, USA}
\affiliation{Joint Quantum Institute, NIST/University of Maryland, College Park, Maryland 20742, USA}
\affiliation{Institute for Research in Electronics and Applied Physics, University of Maryland, College Park, MD 20742, USA}

\begin{abstract}
Synthetic quantum Hall bilayer (SQHB), realized by optically driven monolayer graphene in the quantum Hall regime, provides a flexible platform for engineering quantum Hall phases as discussed in [Phys. Rev. Lett. 119, 247403]. The coherent driving which couples two Landau levels mimicks an effective tunneling between synthetic layers. The tunneling strength, the effective Zeeman coupling, and two-body interaction matrix elements are tunable by varying the driving frequency and the driving strength. Using infinite density matrix renormalization group (iDMRG) techniques combined with exact diagonalization (ED), we show that the system exhibits a non-abelian bilayer Fibonacci phase at filling fraction $\nu = 2/3$. Moreover, at integer filling $\nu = 1$, the SQHB exhibits quantum Hall ferromagnetism. Using Hartree-Fock theory and exact diagonalization, we show that excitations of the quantum Hall ferromagnet are topological textures known as skyrmions.  
\end{abstract}

\maketitle

\section{Introduction}


Fractional quantum Hall (FQH) phases are paradigm examples of topological order, providing the rich physics associated with anyonic statistics \citep{leinaas1977theory,halperin1984statistics,arovas2002fractional}. Moreover, non-abelian anyon statistics \citep{willett1987observation,moore1991nonabelions,greiter1991paired,
nayak19962n,read2000paired} have been shown to be a powerful resource for performing topological quantum computation \citep{sarma2015majorana,vaezi2014superconducting,RevModPhys.80.1083}. Currently, there is intense interest in the realization of FQH states in the multi-component systems \citep{halperin1983theory,haldane1988spin,wen2000continuous,barkeshli2011bilayer}. In contrast to the single component system, the multi-component FQH system with extra degree of freedom enables wider tunablity and exhibits a richer quantum phase diagram. Several non-abelian FQH phases have been proposed for bilayer FQH systems, including the Moore-Read state at filling $\nu=1/2$ \citep{zhu2016fractional}, inter- and intralayer Pfaffian states at filling $\nu=2/3$ \citep{PhysRevB.82.233301,geraedts2015competing}, and bilayer Fibonacci state at filling $\nu=2/3$ \citep{vaezi2014fibonacci,liu2015non}. 
 
In addition to these topological order states, the multi-component quantum Hall system may also exhibit synthetic quantum Hall ferromagnetism. In such ferromagnet, all electrons spontaneously align their (iso-)spin  in order to minimize the Coulomb exchange interaction, while their kinetic energy is quenched into highly degenerate Landau levels (LLs). Adding an additional particle to the ferromagnet triggers a skyrmion excitation, which is characterized by a winding of the magnetization. Skyrmion excitations have been the subject of many theoretical \citep{PhysRevB.47.16419,PhysRevB.50.11018,fertig1997hartree,PhysRevB.51.5138} and experimental studies \citep{barrett1995optically,schmeller1995evidence,aifer1996evidence}. 

It has been shown that the monolayer graphene coupled to a light field enables flexible control on the quantum level \citep{grass2018optical, ghazaryan2017light, PhysRevB.79.081406}. For example, optical driving can be used to induce topologically non-trivial band structure through Floquet mechanism \citep{cayssol2013floquet, PhysRevB.79.081406}. So far, Floquet topological insulators have mainly been studied from the perspective of single-particle physics, but more recently, it has also been proposed to modify effective interaction terms via optical driving \citep{ghazaryan2017light}. This paves the way to the optical engineering of FQH phases. Specifically, when a classical light field couples to two LLs near resonance, the optical transitions between the two Landau levels mimick an effective tunneling between two synthetic ``layers'', so the system can be interpreted as a synthetic quantum Hall bilayer (SQHB). In contrast to real bilayers, the tunneling strength in the SQHB is freely tunable via the laser intensity. The detuning of the coupling can be used to adjust the chemical potential of the two synthetic layers.

\begin{figure}[t]
\includegraphics[width=0.8\columnwidth]{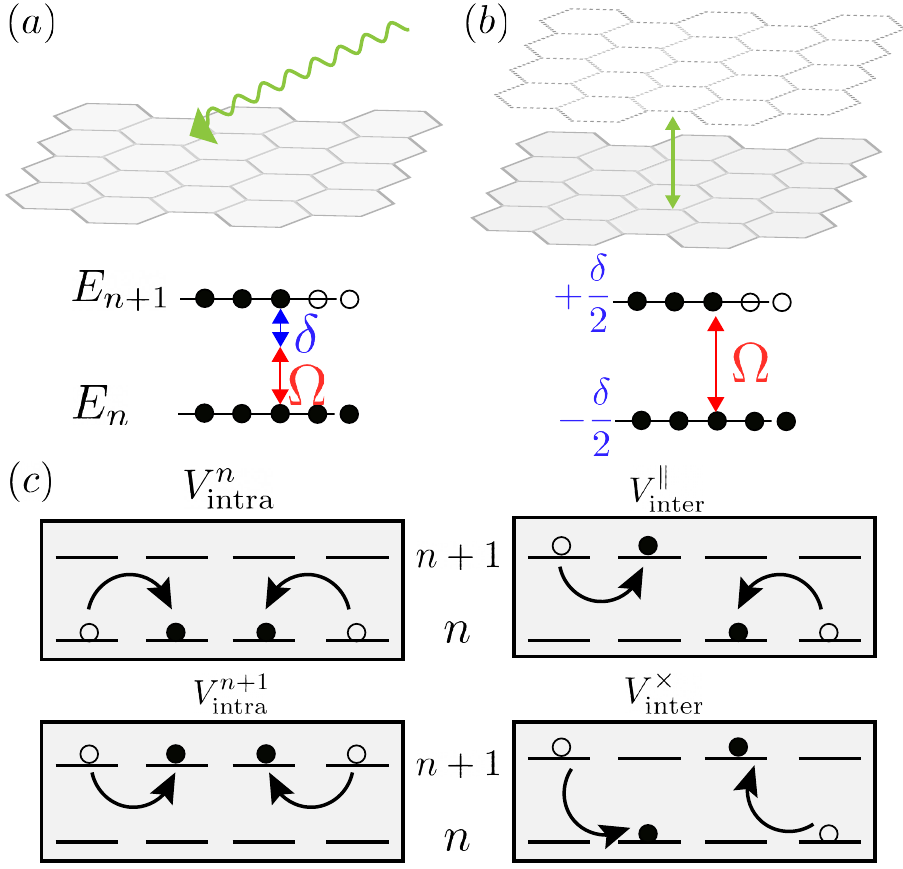}
 \caption{{\bf Illustration of the SQHB system.} (a) A monolayer graphene in the quantum Hall regime is driven by light with Rabi frequency $\Omega$ and detunning $\delta$. (b) In the rotating frame, the system effectively becomes a quantum Hall bilayer. The tunneling strength is given by the Rabi frequency and the energy difference between the two layers is determined by the laser detunning $\delta$. (c) The effective Coulomb interaction in the SQHB picture: $V^n_{\rm intra}$ and $V^{n+1}_{\rm intra}$ are the intra-layer interaction that scatter electrons in the same layer. $V^{\parallel}_{\rm inter}$ is the inter-layer interaction that preserve the layer index. The interaction $V^{\times}_{\rm inter}$ exchanges the layer index of the two electrons during the scattering process. Such a process is absent in the usual quantum Hall bilayer.}
 \label{System}
 \end{figure}
 
One particularly intriguing case is when the first LL ($LL_1$) and the second LL ($LL_2$) are coupled ($LL_1-LL_2$). Then, the repulsion between singlet pairs becomes small. To some extent, these interactions resemble a hollow-core model, that is, an interaction model based on Haldane pseudopotentials $V_m$ \citep{haldane1983fractional}, in which $V_1 \neq 0$, but $V_0=0$.  
Generally, such interactions favor the formation of many-body singlet states, and at filling fraction $\nu = 2/3$, the ground state of the hollow-core Hamiltonian has been reported to be a non-abelian phase. Both the interlayer Pfaffian phase \citep{geraedts2015competing}, and the bilayer Fibonacci phase \citep{PhysRevB.82.233301} have been discussed in this context, but the topological phase of the SQHB at $\nu = 2/3$ has remained unclear in the previous study, mainly due to the limitation of small system size accessible to the exact diagonalization (ED) sutdy. Here, by using the infinite density matrix renormaliztion group (iDMRG)\citep{mcculloch2008infinite,zaletel2013topological,zaletel2015infinite} along with ED, we identify the non-abelian phase of the $LL_1-LL_2$ synthetic bilayer system to be the bilayer Fibonacci phase. With this, the SQHB becomes an interesting environment for topological quantum computing.

The striking FQH behavior of the $LL_1-LL_2$ SQHB is a consequence of the peculiar shape of its pseudopotentials. The behavior is very observed in a system where $LL_0$ and $LL_1$ are coupled. As we show in this paper, the qualitative change of interactions in $LL_0-LL_1$ or $LL_1-LL2$ bilayers can also be observed at integer filling $\nu=1$, although interactions typically play a much smaller role in the integer quantum Hall regime. Specifically, we show that the $LL_0-LL_1$ bilayer exhibits synthetic quantum Hall ferromagnetism at $\nu=1$, whereas the $LL_1-LL_2$ bilayer does not, due to its tendency towards singlet formation. 

The synthetic ferromagnetic behavior can also be controlled by the laser detuning: It acts as an effective Zeeman term which lifts the (iso-)spin degeneracy and competing with the ferromagnetic exchange energy whose scale is given by the strength of the Coulomb interaction. When the ferromagnetic exchange interaction dominates over the Zeeman energy, the addition of one particle leads to a spin flip of many particles in order to keep neighboring spins almost aligned with each other. This collective spin flip leads to a winding texture, which is known as a skyrmion. Using Hartree-Fock mean-field theory and exact diagonalization, we show that the $LL_0-LL_1$ SQHB system exhibits such skyrmion excitations, whereas the $LL_1-LL_2$ system does not.

The paper is organized as follows: In Sec. II, we review the formalism of the graphene quantum Hall state and the SQHB. In Sec. III, we provide the detail numerical evidence showing that the SQHB with filling fraction $\nu = 2/3$ is a bilayer Fibonacci phase. In Sec. IV, the iso-spin texture excitation in the quantum Hall ferromagnetic regime is discussed. Finally,  in section V, we summarize our results.

\section{Synthetic Bilayer Graphene System}
In this section, we describe the SQHB system, that is, a single-layer quantum Hall system in which a synthetic bilayer degree of freedom is induced by a laser-coupling between Landau levels. Such a system can be realized in monolayer graphene under a strong magnetic field \citep{RevModPhys.83.1193}. We assume that both the electronic spin and valley degree of freedoms are fully polarized. In the quantum Hall regime, the single-particle eigenstates in graphene are given by spinors of the form 
\begin{align}
\psi_{\sigma, n, m} = 
\begin{bmatrix}
C_n^-\ket{n-1, m} \\
C_n^+\sigma \ket{n, m}
\end{bmatrix},
\end{align}
where $C_n^\pm = \sqrt{\frac{1\pm \delta_{0,n}}{2}}$. The quantum number $\lambda = \pm 1$  labels the states of positive and negative energy, respectively. The kets $\ket{n,m}$ denote the eigenstates of a non-relativistic quantum Hall system, with $\langle {\bf r} | n,m \rangle$ being the Landau level (LL) wave function of the $n$th LL with orbital quantum number $m$. In the symmetric gauge, the orbital quantum number $m$ denotes the angular momentum. In the Landau gauge, it represents the momentum which is conserved along one spatial direction. The single-particle energy of the state $\psi_{\sigma, n, m}$ is given by 
\begin{align}
E_\sigma = \sigma \frac{\hbar v_F}{l_B}\sqrt{2n},
\end{align}
where $l_B = \sqrt{\hbar/eB}$ is the magnetic length, $B$ is the magnetic field strength and $v_F$ is the Fermi velocity. In the following, we only consider the positive energy part and, therefore, drop the subscript $\sigma$ for simplicity.

Unlike the non-relativistic LL spectrum, the relativistic energy spectrum is not quantized at equally spaced values in the graphene quantum Hall system. We can thus selectively couple two distinct LLs via a mono-chromatic laser with frequency $\omega_L$, according to the usual selection rule $|n| \leftrightarrow |n \pm 1|$. These two laser-coupled LLs represent the two ``layers'' of our synthetic bilayer quantum Hall system.

In the following, we consider a coupling between a (partially) filled LL and an empty $(n+1)$th LL. We assume that the driving laser is a plane wave such that the coupling is non-vanishing only between states with the same orbital quantum number $m$. Other selection rules are possible if the light is designed to have orbital angular momentum \citep{grass2018optical, PhysRevB.95.235439}. Under the rotating wave approximation (RWA), the Hamiltonian of the synthetic bilayer system  is given by \citep{ghazaryan2017light}
\begin{align}
&H = H_0 + H_{int},\notag \\ 
&H_0 = \sum_m \left( -\delta\tau^z_{n,m} + \Omega \tau^x_{n,m} \right),\notag \\
&H_{int} = \sum_{n_1+n_2=n_3+n_4}\sum_{ \{m\} }V^{n_1,n_2,n_3,n_4}_{m_1,m_2,m_3,m_4} c^\dagger_{n_1, m_1}c^\dagger_{n_2, m_2}c_{n_3, m_3}c_{n_4, m_4},
\label{H_0}
\end{align}
where $c_{n,m} $ and $c^\dg_{n,m}$ are the annihilation and creation operators in the $n$th Landau level with angular momentum quantum number $m$,  $\delta = E_{n+1} - E_{n} - \omega_L$ is the detuning, and $\Omega$ is the Rabi frequency. The iso-spin operators are given by $\tau^z_{n,m} = c^\dg_{n,m}c_{n,m} - c^\dg_{n+1,m}c_{n+1,m}$ and $\tau^x_{n,m} = c^\dg_{n,m}c_{n+1,m} + c^\dg_{n+1,m}c_{n,m}$. The first term in $H_0$ corresponds to an effective Zeeman coupling for the quantum Hall system with spin degree of freedom and the second term corresponds to the tunneling in the bilayer quantum Hall system. The interaction matrix elements $V^{n_1,n_2,n_3,n_4}_{m_1,m_2,m_3,m_4}$ are for Coulomb scattering of a pair of electrons in Landau orbitals $\{n_1,m_1\}$ and $\{n_2,m_2\}$ to orbitals $\{n_3,m_3\}$ and $\{n_4,m_4\}$, but the sum over $n_i$ is restricted to $n_1+n_2=n_3+n_4$ by the RWA.

In a conventional bilayer system, the spatial overlap between single-particle states in different layers is negligible, and thus, Coulomb terms which would scatter an electron from one layer into the other do not play a role. With this, the layer index of each particle is conserved in the scattering term. For the synthetic bilayer, however, the situation is different, as two particles can exchange their individual indices. This leads to the four different types of scattering process which are shown in \ref{System}(c): The intra-layer interaction in the $n$th LL and $(n+1)$th LL are $V^{n}_{\rm intra} = \sum_{ \{m\} }V^{n,n,n,n}_{m_1,m_2,m_3,m_4}c^\dagger_{n, m_1}c^\dagger_{n, m_2}c_{n, m_3}c_{n, m_4}$, and $V^{n+1}_{\rm intra}$ respectively. For a pair of electrons in different layers, there are two types of inter-layer interactions: the standard process which keeps the electrons in their layer is denoted by $V^{\parallel}_{\rm inter} = 2 \sum_{ \{m\} }V^{n,n+1,n+1,n}_{m_1,m_2,m_3,m_4}c^\dagger_{n, m_1}c^\dagger_{n+1, m_2}c_{n+1, m_3}c_{n, m_4}$. In addition to this, the SQHB allows for an exchange interaction in which the layer index is changed, i.e. $V^{\times}_{\rm inter} = 2 \sum_{ \{m\} }V^{n,n+1,n,n+1}_{m_1,m_2,m_3,m_4}c^\dagger_{n, m_1}c^\dagger_{n+1, m_2}c_{n, m_3}c_{n+1, m_4}$. 

We may expand these different scattering potentials in terms of Haldane pseudopotentials, $V^n_m, V^{n+1}_m, V^\parallel_m, V^\times_m$. As shown in Ref. \citep{ghazaryan2017light}, the
scattering of interlayer singlets is described by the pseudopotentials $V^\parallel_m-V^\times_m$, whereas the scattering of interlayer triplets is given by $V^\parallel_m+V^\times_m$.
Noting that a symmetric (antisymmetric) layer configuration has to be combined with an antisymmetric (symmetric) spatial wave function, the interlayer scattering at even (odd) values of $m$ is given by the pseudopotentials for interlayer singlets (triplets):
\begin{align}
\label{vinter}
  V_m^{\rm inter} = \begin{cases}
 V_m^{\|} + V_m^\times &\text{if $m$ is odd,}\\
 V_m^{\|} - V_m^\times &\text{if $m$ is even.}
\end{cases}
\end{align}

The form of the interlayer potential highlights the role which is played by the exchange interaction $V^\times_m$: While suppressing the scattering at $m=0$, it enhances interactions at $m=1$. As shown in Ref. \citep{ghazaryan2017light}, the strength of this effect depends crucially on the Landau levels which are coupled: When the zeroth and the first Landau levels are coupled ($LL_0-LL_1$), the effect of the exchange interaction is only quantitative (in the sense that $V_0^{\rm inter}$ remains the strongest interlayer interaction channel). In contrast, when the first and the second Landau level are coupled ($LL_1-LL_2$), a qualitative change of $V_m^{\rm inter}$ is seen. In this case, $V^{\rm inter}_1>V^{\rm inter}_0$, that is, the first Haldane pseudopotential dominates the interlayer interaction. Therefore, the synthetic bilayer with $LL_1-LL_2$ coupling has a strong tendency to form spin singlet phases.

\section{Bilayer Fibonacci phase}
A huge variety of spin singlet phases have been discussed for bilayers at filling fraction $\nu = \frac{2}{3}$. These phases include Abelian composite fermion and Halperin phases, and also non-Abelian phases such as bilayer Fibonacci state and interlayer-Pfaffian state \cite{PhysRevB.82.233301,geraedts2015competing,vaezi2014fibonacci,liu2015non}. Strikingly, in the SQHB non-zero overlaps have been reported for these non-abelian phases, but a clear identification of the phase has remained a challenge. Below, we provide a variety of numerical evidences which demonstrate that the SQHB exhibits the bilayer Fibonacci phase. Specifically, we compute various characteristics of topological phases, including entanglement spectra, entanglement entropy, ground state degeneracies, using the large scale infinite density matrix renormalization group (iDMRG) algorithm on infinite cylinder geometry, as well as ED in a spherical geometry. 

 Before presenting the numerical evidences for the Fibonacci phase, let us briefly discuss the role played by the parameters in the single-particle Hamiltonian. The single-particle orbitals are superpositions of the two synthetic layers, and for $\Omega \gg \delta$, the orbitals are simply the symmetric and anti-symmetric combinations. The energy splitting between the two states are of the order of $2\Omega$, with the anti-symmetric orbits being the lower manifold. If this single-particle gap becomes large as compared to the interaction energy, i.e. $\Omega \gg \frac{e^2}{2\epsilon l_B}$, we can treat the system as a single layer quantum Hall system. In this case, the ground state at $\nu=2/3$ is the hole-conjugate of the $\nu = \frac{1}{3}$ Laughlin state. The system undergoes a phase transition as the Rabi frequency $\Omega$ is decreased. In the weak coupling regime, the ground state forms a layer singlet state \citep{ghazaryan2017light}, which is identified as the Fibonacci phase below.
 
For the bilayer Fibonacci phase, there are six topologically distinct types of quasi-particles. Three of them are abelian quasi-particle denoted by $\Phi_n$, where $n = 0, 1, 2$. The abelian quasi-particles follow the fusion rule $\Phi_a \times \Phi_b = \Phi_{(a+b)\%3}$.  The $\Phi_0$ sector corresponds to the vacuum sector $\mathcal{I}$ since it satisfies $\Phi_0 \times \Phi_n = \Phi_n$. On the other hand, there is a "Fibonacci" quasi-particle $\tau$ which satisfies the fusion rule $\tau \times \tau = 1 + \tau$. The braiding statistics of the Fibonacci anyon allows for universal topological quantum computation. The rest of  two quasi-particles are can be described by $\Phi_a\tau$ with $a = 1,2$  \citep{vaezi2014fibonacci,liu2015non}. 

The evidences for characterizing the bilayer Fibonacci phase are the following: (1) We perform the adiabatic continuation (AC) to show that the ground state on the sphere is in the same class as the bilayer Fibonacci phase.  (2) We obtain two topologically distinct degenerate ground states $\ket{\psi_1}$ and $\ket{\psi_2}$ on a infinite cylinder. Combined with the center-of-mass translation, this leads to a six-fold ground state degeneracy. (3) The counting of edge states is done within the orbital entanglement spectrum obtained from these two ground states, and it matches the counting expected for the bilayer Fibonacci phase. (4) By calculating the difference of the entanglement entropy and the momentum polarization between the two ground states $\ket{\psi_1}$ and $\ket{\psi_2}$, we obtain the topological entanglement entropy and the topological spin of the non-abelian anyon. Their values are consistent with the bilayer Fibonacci phase.  

\subsection{Adiabatic Continuation on the Spherical Geometry}
The finite size limitation of the ED calculation make it challenging to extract useful topological information.
However, since topological phase transitions require the closing of the energy gap, it is possible to test the topological behavior by adiabatically deforming the system Hamiltonian into a simpler model Hamiltonian for which the topological phase is known. We adiabatically change the electron interaction by interpolating between the Coulomb interaction of the synthetic graphene bilayer and a hollow-core model, that is, an interaction Hamiltonian with the  interlayer pseudopotential $V_1^{\rm inter}$ being the only-non-zero pseudopotential. Such a model has been shown to support the bilayer Fibonacci phase \cite{liu2015non}. The Hamiltonian which interpolates between Coulomb interaction and hollow-core model is given by
\begin{align}
H_\lambda = (1-\lambda)H_{int} + \lambda \hat V^{{\rm inter}}_1,
\label{H_AC}
\end{align}
where $0\leq \lambda \leq 1$. $H_{int}$ is the Coulomb interaction of the synthetic bilayer graphene system, and  $\hat V^{{\rm inter}}_1$ is the interaction term generated by a  interlayer Haldane pseudo-potential model with $V_m^{\rm{inter}}=\delta_{m,1}$ . If the ground state of the synthetic bilayer graphene and the $\hat V^{{\rm inter}}_1$ interaction are in the same universality class, the ground state wave function and the ground state energy should change smoothly when the parameter $\lambda$ in the Hamiltonian in Eq.~\eqref{H_AC} varies adiabatically, and the gap above the ground state should not close.

\begin{figure}[h]
 \includegraphics[width=1\columnwidth]{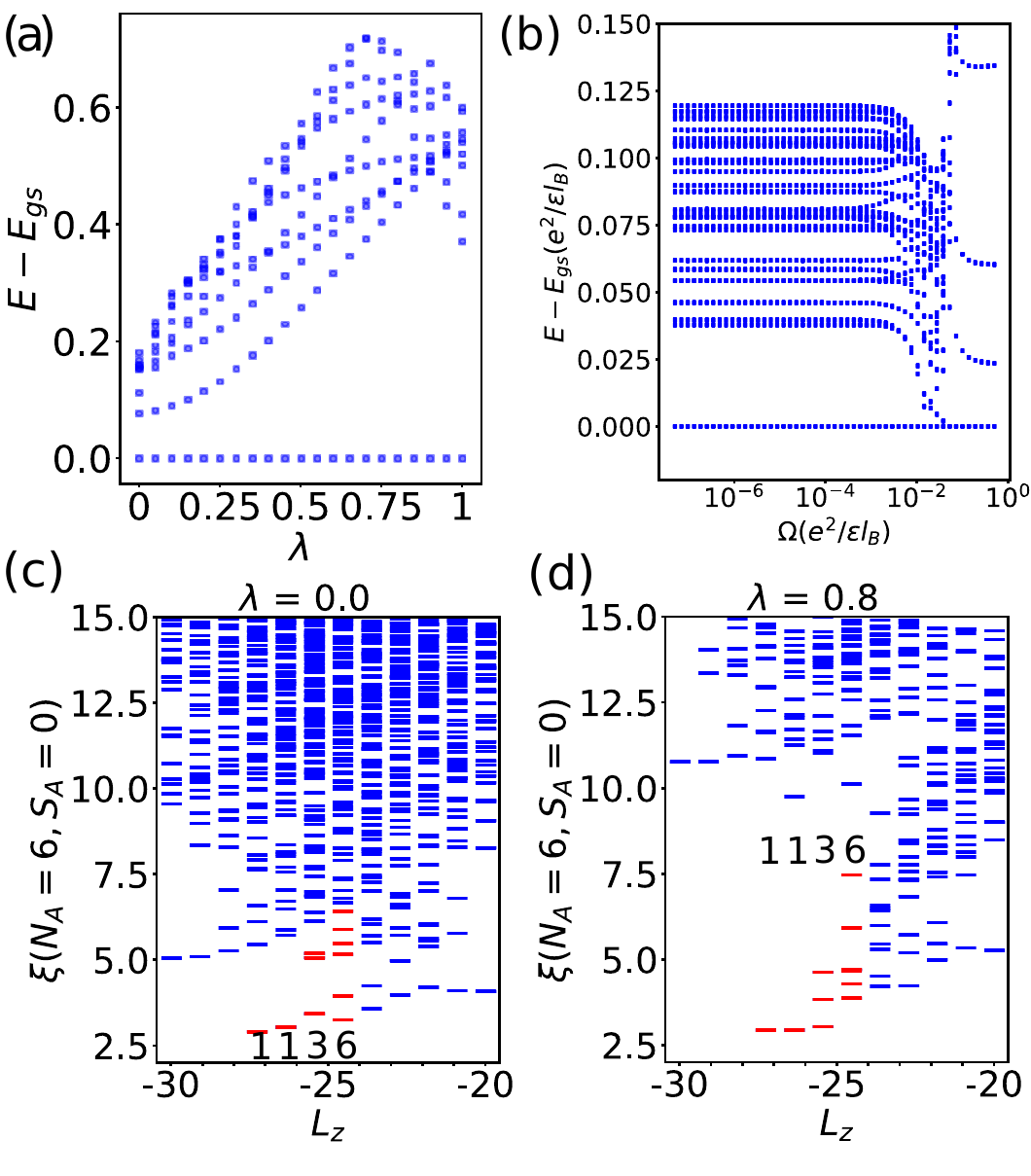}
  \caption{ {\bf Adiabatic continuation via exact diagonalization}  (a) Energy spectrum in the process of adiabatic continuation for $N_e = 12$, $N_\phi = 15$. The energy gap remains open during the adiabatic continuation process. This indicates that the ground state wavefunction and bilayer Fibonacci phase are in the same class. (b) Energy spectrum  for $N_e = 8$, $N_\phi = 10$ and $\lambda = 0$ as function of the Rabi frequency $\Omega$. The system undergoes a phase transition to the particle-hole conjugate of Laughlin state when $\Omega \gg 1$. (c) and (d) show the orbital cut entanglement spectrum for $N_e = 12$, $N_\phi = 15$,with $\lambda = 0$ and $\lambda = 0.8$ respectively.   }
\label{ED_AC}
\end{figure}

To access the energy spectrum of the Hamiltonian in Eq.~\eqref{H_AC}, the ED calculation is performed in the spherical geometry. The number of electrons $N_e$ and total number of quantum fluxes $N_\phi$ are related by $N_\phi = \frac{1}{\nu}N_e -S$, where $S$ is the shift of fractional QH state on sphere. For the bilayer Fibonacci phase, the shift $S = 3$. We start with infinitesimal Rabi frequency $\Omega$ and zero detuning $\delta$ and therefore occupation of each synthetic layer is conserved. Thus, we can to examine the entanglement spectrum with a fixed total layer polarization (or total pseudospin). The stability upon increasing the Rabi frequency $\Omega$ to a finite value in the weak coupling regime  is shown in the fig. \ref{ED_AC}(b). The energy gap remains open under small perturbation of $\Omega$. 

The energy spectrum of the adiabatic continuation with $N_e = 12$ is shown in Fig. \ref{ED_AC}(a). The energy gap remains open in the process of the adiabatic continuation. This suggests that the ground state of the synthetic bilayer graphene is in the phase defined by $\hat V_1^{\rm inter}$, i.e. the bilayer Fibonacci phase. 

Fig. \ref{ED_AC}(c) and (d) show the orbital cut entanglement spectrum (OES) for zero and non-zero value of $\lambda$. The low energy part of the ES corresponds to the degeneracy of the edge excitation with angular momentum $L_z$ relative to the ground state \citep{PhysRevLett.101.010504}. The counting of the edge excitation allows us to determine the topological order of a wave function. However, finite-size effects close the entanglement gap in most angular momentum sectors, which makes the correct counting difficult. However, we are able to verify that the counting for the SQHB [see panel Fig.\ref{ED_AC}(c)] is compatible with the counting 1,1,3,6 which is characteristic for the Fibonacci phase, and which we obtain when $\lambda$ is chosen to be close to 1 [see panel Fig.\ref{ED_AC}(d)]. 

\begin{figure}[b]
 \includegraphics[width=1\columnwidth]{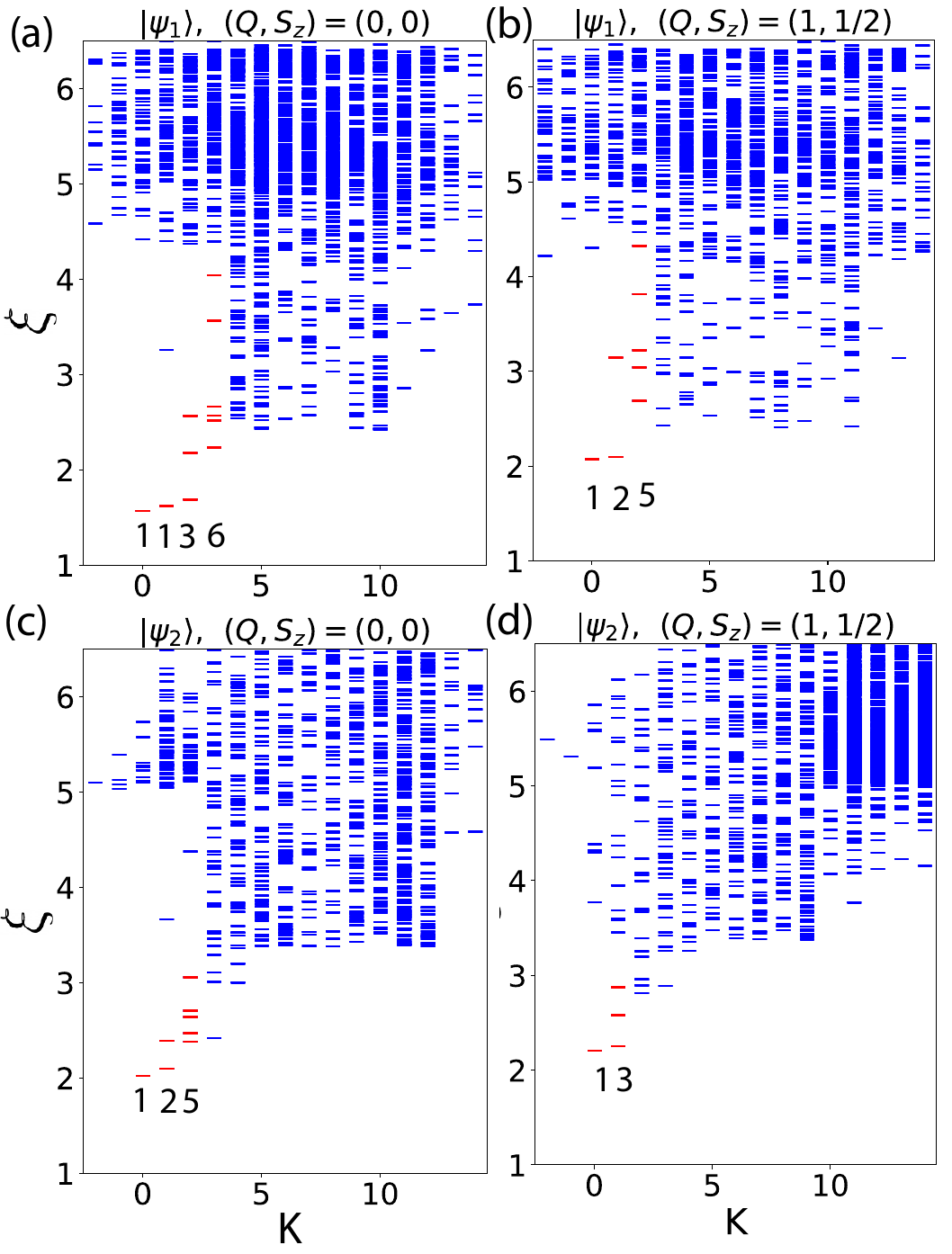}
  \caption{ {\bf Entanglement spectrum via iDMRG simulation.} Orbital entanglement spectra for the two degenerate ground states at circumference $L = 30 l_B$. The maximum bond dimension $\chi_{max} = 15000$. Two distinct orbital cuts are presented for each ground state. The low-lying spectra agree with the CFT prediction as described in the main text. \label{iDMRG_ES}}
\end{figure}

\subsection{Ground State Degeneracy and Entanglement Spectrum}
On topological non-trivial geometries, such as the torus, non-Abelian phases exhibit characteristic ground state degeneracies. The same degeneracies as for the torus can also be observed for infinite cylinders, for which we have computed the ground states via infinite DMRG (iDMRG) \citep{zaletel2013topological}. 

We perform the iDMRG simulation and obtain two orthogonal states $\ket{\psi_1}$ and $\ket{\psi_2}$. The two wave functions are nearly degenerate. Along with the center-of-mass translation, we have an (at least) six-fold degenerate ground state on the infinite cylinder. It should be noted, though, that this is as a lower bound on the degeneracy, since there is no systematic way to guarantee that iDMRG finds all degenerate ground states. In order to further confirm our result, we initialize the infinite MPS ansatz with different configuration for perimeter of the infinite cylinder $4l_B \leq L \leq 10l_B$. After a few DMRG sweeps with moderate bond dimention (around 450), the wave function always converges to two orthogonal states. 

In the following section, we examine the topological properties of the two wave functions by calculating the orbital entanglement spectrum (OES), topological entanglement entropy (TEE), and the topological spin. In order to resolve the entanglement spectrum in different total spin sector, we assume that the Rabi frequency $\Omega$ is infinitesimal and, therefore, the total pseudo-spin $S_z = N_{\uparrow} - N_{\downarrow}$ is a good quantum number, where $N_{\uparrow}$ and $N_{\downarrow}$ denote the number of electrons in the $nth$ and $(n+1)th$ LLs, respectively. Fig.~\ref{iDMRG_ES} shows the entanglement spectrum of the ground state $\ket{\psi_1}$ and $\ket{\psi_2}$. The level counting of the OES can be obtained using the thin torus patterns and generalized exclusion rules \citep{liu2015non}. For the ground state $\ket{\psi_1}$, the counting is $1, 1, 3$ when the partition has even charge $Q$ and it is $1, 2, 5$ when the partition has odd charge $Q$, as shown in Fig.~\ref{iDMRG_ES} (a) and (b). For state $\ket{\psi_2}$, the spectrum follows $1, 2, 5$ for even $Q$ and $1, 3, 6$ for odd $Q$ as shown in Fig.~\ref{iDMRG_ES} (c) and (d).

\begin{figure}[b]
 \includegraphics[width=1\columnwidth]{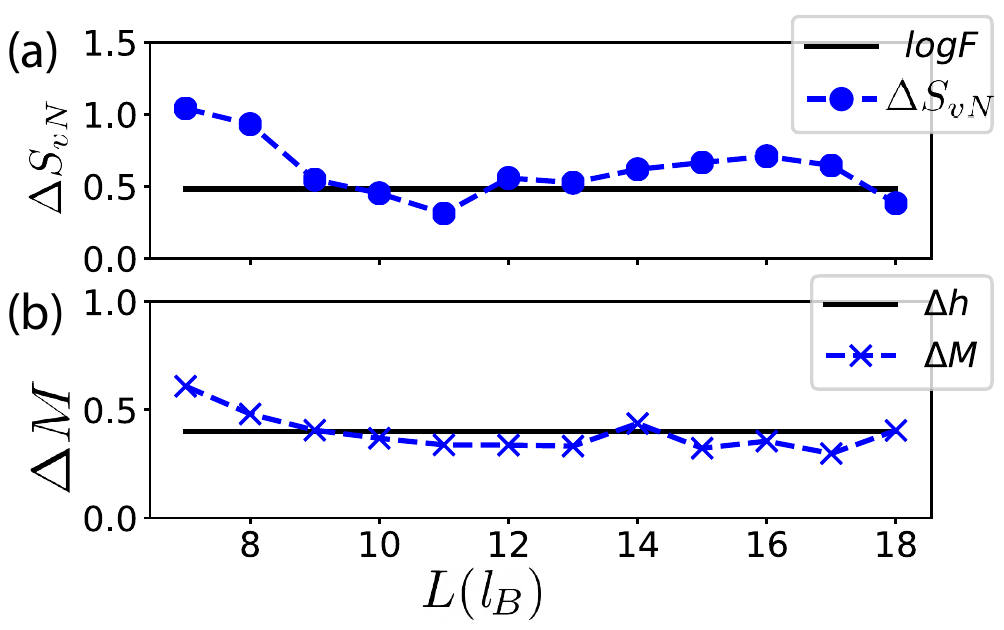}
  \caption{ iDMRG simulation of the differences in (a) entanglement entropy ($\Delta S_{vN}$) and (b)
momentum polarization ($\Delta M$) for the two degenerate states as a function
of circumference ($L$).  The black lines show the theoretical value of $\Delta S_{\bf vN} = S^\tau_{\rm vN} - S^{\mathcal{I}}_{\rm vN}$ and $\Delta M = h_{\tau} - h_{\mathcal{I}}$ for bilayer Fibonacci phase as described in the main text. }
\label{iDMRG_E_TEE_S}
\end{figure}

\subsection{Topological Entanglement Entropy and Topological Spin}
The entanglement entropy $S_{{\rm vN}}^a$ of a FQH system with an anyon type $a$ on the infinite cylinder scales as 
\begin{align}
S_{{\rm vN}}^a = \alpha L - \gamma^a,
\end{align}
where $\alpha$ is a non-universal constant, $L$ is the perimeter of the cylinder. By $\gamma^a$, we denote the topological entanglement entropy of the anyon type $a$, which is related to the quantum dimension $d_a$ of the anyon $a$, and the total quantum dimension $\mathcal{D}$ of the topological phase via the relation  $\gamma^a = \log(\mathcal{D}/d_a)$.

We calculate the difference of the entanglement entropy between the two ground states. Since the quantum dimension of the two topological sectors are given by $d_{ \mathcal{I} } = 1$ and $d_{\tau} = F =  \frac{1+\sqrt{5}}{2}$, which is the golden ratio, we have
\begin{align}
\Delta S_{{\rm vN}} = S_{{\rm vN}}^\tau - S_{{\rm vN}}^{\mathcal{I}} = \log(d_\tau / d_{\mathcal{I} }) \approx 0.48.
\end{align}

Figure ~\ref{iDMRG_E_TEE_S}(a) shows the result of $\Delta S_{{\rm vN}}$. Due to finite size effects and the truncation error of the bond dimension, the data exhibits a significant systematic error. Although we do not determine the quantum dimension unambiguously, the results are still consistent with bilayer Fibonacci phase $d_{\tau}  = F$. 

The momentum polarization $M^a $ computes the Berry phase in the process of twisting the left half of the infinite cylinder. It is defined as $M^a = {\rm Tr}(\rho^a_L K)$ where $\rho_L$ is the density matrix of the anyon type $a$ in the left half of the infinite cylinder and $K$ is the momentum operator on the cylinder. The momentum polarization is related to three topological invariants: the shift $S$, the topological spin $h$ and the central charge $c$ \citep{zaletel2015infinite}:
\begin{align}
M^a = -\frac{\nu S L^2}{(4\pi)^2} + h_a - \frac{c}{24}~{\rm (mod~1)},
\end{align}
where $L$ is the perimeter of the infinite cylinder and $\nu$ is the filling fraction. The difference of the momentum polarization between the two topological sectors is of the form 
\begin{align}
\Delta M = M^\tau - M^{\mathcal{I}} = h_{\tau} - h_{\mathcal{I}}~{\rm (mod~1)}.
\end{align}
The different of topological spin is given by $h_\tau - h_\mathcal{I} = 2/5$. Fig.~\ref{iDMRG_E_TEE_S}(b) shows our calculation of the momentum polarization, which is consistent with the CFT prediction \citep{vaezi2014fibonacci}. 

\section{Spin textures at integer filling}
The results presented in the previous section show that the synthetic quantum Hall bilayer is a promising candidate for realizing intriguing non-Abelian phases of matter.
The crucial ingredient which gives rise to the non-Abelian behavior is the strong enhancement of the interlayer pseudpotential $V_1^{\rm inter}$, when the synthetic layers are given by $LL_1$ and $LL_2$. It is interesting to further investigate the role of these excotic interactions in the synthetic bilayer. This section takes a look onto the integer quantum Hall regime, which in comparison to the regime of fractional filling factors, is technically less difficult to realize. In this context, we will focus on the charged excitation of the integer quantum Hall system, in which interactions may give rise to interesting spin structures.

In the integer quantum Hall regime, interactions play a role when the gap between two Landau levels becomes comparable to the Coulomb interaction energy. In the literature, such a situation has first been considered for systems where the two Landau levels are given by manifolds of opposite spin, separated by the Zeeman gap \citep{PhysRevB.47.16419}. It has been shown that, if ferromagnetic exchange interactions overweigh the single-particle gap, then the elementary charged excitation will be a collective excitation of many electrons occupying  the upper Zeeman manifold. This results in a spin texture which slowly winds around when going from the center to the edge of the system, known as a skyrmion. Similar pseudospin textures have been discussed for bilayer quantum Hall systems \citep{PhysRevB.51.5138}. Here, we will investigate the spin textures in the synthetic quantum Hall bilayer using exact diagonalization and mean-field techniques.

The single-particle part of the synthetic bilayer system is described in Eq.\eqref{H_0},
\begin{align}
H_0 = \sum_{m} -\delta \tau^z_{n,m} + \Omega \tau^x_{n,m} = \sum_{m}\omega_g \tilde{\tau}^z_{n,m},
\end{align}
where the effective Zeeman energy $\omega_g = \sqrt{\delta^2 + \Omega^2}$, $\tilde{\tau}^z_{n,m} = \cos\theta \tau^z_{n,m} + \sin\theta \tau^x_{n,m}$ and $\theta = \tan^{-1} \frac{-    \Omega}{\delta}$. As in usual spin systems, the Zeeman energy tends to polarize the electrons (at finite $\Omega$ in a dressed state), and the Coulomb interaction competes with it. However, 
in contrast to a physical bilayer, the effect of Coulomb interaction is significantly different in the synthetic bilayer. In the following, we present our results from a mean-field approach and and from an exact numerical treatment.

\subsection{Mean field approach}

In the case of non-relativistic spin-1/2 quantum Hall systems, it has been shown that skyrmionic spin textures are obtained within a mean-field description \citep{PhysRevB.50.11018}. This approximation replaces the fourth-order operator products in $H_{int}$ by second-order products, \[c^\dagger_1 c^\dagger_2 c_3c_4 = \langle c^\dagger_1 c_4 \rangle c^\dagger_2 c_3 - \langle c^\dagger_2 c_4 \rangle c^\dagger_1 c_3 + \langle c^\dagger_2 c_3 \rangle c^\dagger_1 c_4 - \langle c^\dagger_1 c_3 \rangle c^\dagger_2 c_4.\] By applying this approximation, one obtains a quadratic Hamiltonian, and truncating to $M$ orbitals per Landau level, the system is described by a $2M\times2M$ matrix. To ease the notation, we define $a_i \equiv c_{n+1,m_i}$, $b_i \equiv c_{n,m_i}$,  $a_i^\dagger \equiv c^\dagger_{n+1,m_i}$, $b_i^\dagger \equiv c^\dagger_{n,m_i}$. Further, we denote interaction matrix elements by $V_{1234}^{x_1,x_2,x_3,x_4}$, with $x_i =\{a,b\}$, and the subscript being a short-hand notation for the orbitals $m_i$ of the scattered electrons. We distinguish between three contributions to the mean-field interactions, $V_{\rm HF} = V_{\rm H} - V_{\rm X} + V_{\rm bg}$, which read:
\begin{align}
 V_{\rm H} =& \sum_{\{m\}} \Big( V_{1234}^{aaaa} \langle a_2^\dagger a_3 \rangle a_1^\dagger a_4 + V_{1234}^{bbbb} \langle b_2^\dagger b_3 \rangle b_1^\dagger b_4+ \nonumber \\ &
 +V_{1234}^{baab} \langle a_2^\dagger a_3 \rangle b_1^\dagger b_4 + V_{1234}^{abba} \langle b_2^\dagger b_3 \rangle a_1^\dagger a_4+ \nonumber \\ &
 +V_{1234}^{abab} \langle b_2^\dagger a_3 \rangle a_1^\dagger b_4 + V_{1234}^{baba} \langle a_2^\dagger b_3 \rangle b_1^\dagger a_4 \Big),
 \end{align}
being the Hartree potential,
\begin{align}
  V_{\rm X} =& \sum_{\{m\}} \Big( V_{1234}^{aaaa} \langle a_1^\dagger a_3 \rangle a_2^\dagger a_4 + V_{1234}^{bbbb} \langle b_1^\dagger b_3 \rangle b_2^\dagger b_4+ \nonumber \\ &
 +V_{1234}^{baab} \langle b_1^\dagger a_3 \rangle a_2^\dagger b_4 + V_{1234}^{abba} \langle a_1^\dagger b_3 \rangle b_2^\dagger a_4+ \nonumber \\ &
 +V_{1234}^{abab} \langle a_1^\dagger a_3 \rangle b_1^\dagger b_4 + V_{1234}^{baba} \langle b_2^\dagger b_3 \rangle a_1^\dagger a_4 \Big),
\end{align}
being the exchange potential,
\begin{align}
 V_{\rm bg} = - \sum_{\{m\}} \Big( V_{1221}^{abba} a_1^\dagger a_1 + V_{1221}^{abba} b_1^\dagger b_1),
\end{align}
being the potential which stems from a uniform positive background (identical to a completely filled $b$-level). We stress that, in contrast to a spin system or a real bilayer, the interactions are not SU(2) invariant. We also highlight the existence of flipping terms, $V_{1234}^{abab}$ and $V_{1234}^{baba}$, which are not present in spin systems or real bilayers.

The mean-field Hamiltonian is then solved self-consistently: An initial guess for the correlators defines the Hamiltonian $H_{\rm MF} = H_0+V_{\rm HF}$, and the many-body eigenstates of the Hamiltonian define the correlators. Iteratively, this leads to a self-consistent solution. To calculate the correlators from the eigenstates of $H_{\rm MF}$, we note that the mean-field ground state is given by a Slater determinant over the $N$ lowest single=particle levels, where $N$ is the number of electrons. 

Writing the $k$th single-particle level as $|\Psi^k \rangle = \sum_i \big( \alpha^k_i a_i^\dagger + \beta^k_i b_i^\dagger \big) |{\rm vac} \rangle$, the correlators with respect to the Slater determinant over the levels $1\leq k \leq N$ are given by $\langle a_i^\dagger a_j \rangle = \sum_{k=1}^N \alpha^{k*}_i \alpha^k_j$,  $\langle b_i^\dagger b_j \rangle = \sum_{k=1}^N \beta^{k*}_i \beta^k_j$, and $\langle a_i^\dagger b_j \rangle = \sum_{k=1}^N \alpha^{k*}_i \beta^k_j$.

As there are different fix points, the self-consistent solutions will not be independent from the initial guess, and to obtain a skyrmion solution, the  initial guess shall already contain the skyrmionic correlations. As we are going to consider skyrmions with one electron added to the ferromagnetic ground state, the characteristic skyrmion correlations are as follows: If the ferromagnetic ground state is polarized in the $b$-manifold, the skyrmion is characterized by one $a$-particle in the center ($m=0$), and the other particles occupy single-particle states which are superpositions of $b_m^\dagger |{\rm vac}\rangle$ and $a_{m+1}^\dagger |{\rm vac}\rangle$. The skyrmionic correlations are then characterized by non-zero coherences  $\langle a_{m+1}^\dagger b_{m} \rangle$ and $\langle b_{m}^\dagger a_{m+1}\rangle$, and the spin polarization winds from $a$-polarized in the center to $b$-polarized at the edge.
 In contrast, if the ferromagnetic ground state is polarized in the $a$-manifold, the skyrmion has a $b$-particle in the center, its single-particle orbitals are spanned by  $a_m^\dagger |{\rm vac}\rangle$ and $b_{m+1}^\dagger |{\rm vac}\rangle$, and the characteristic coherences are given by $\langle a_{m-1}^\dagger b_{m} \rangle$ and $\langle b_{m}^\dagger a_{m-1}\rangle$. The spin winding then goes from $b$-polarized in the center to $a$-polarized at the edge.
As a side remark, we note that when the skyrmionic coherences are chosen as the only non-zero coherences in the initial guess, as has been done in Ref. \citep{PhysRevB.50.11018}, the Hartree-Fock Hamiltonian decouples into $M$ 2-by-2 matrices. We also note that in all cases, the occupations are constrained by the number of electrons, $\langle \sum_i (a_i^\dagger a_i + b_i^\dagger b_i)\rangle =N$, which we define such that $N=M+1$. These occupations also provide a natural bound for any of the coherences $\langle X^\dagger Y \rangle \leq \sqrt{n_X n_Y}$, where $X,Y \in \{ a_i, b_i \}$ and $n_X=\langle X^\dagger X\rangle$.

\paragraph{Layer occupation.}
\begin{figure}[h]
 \includegraphics[width=1\columnwidth]{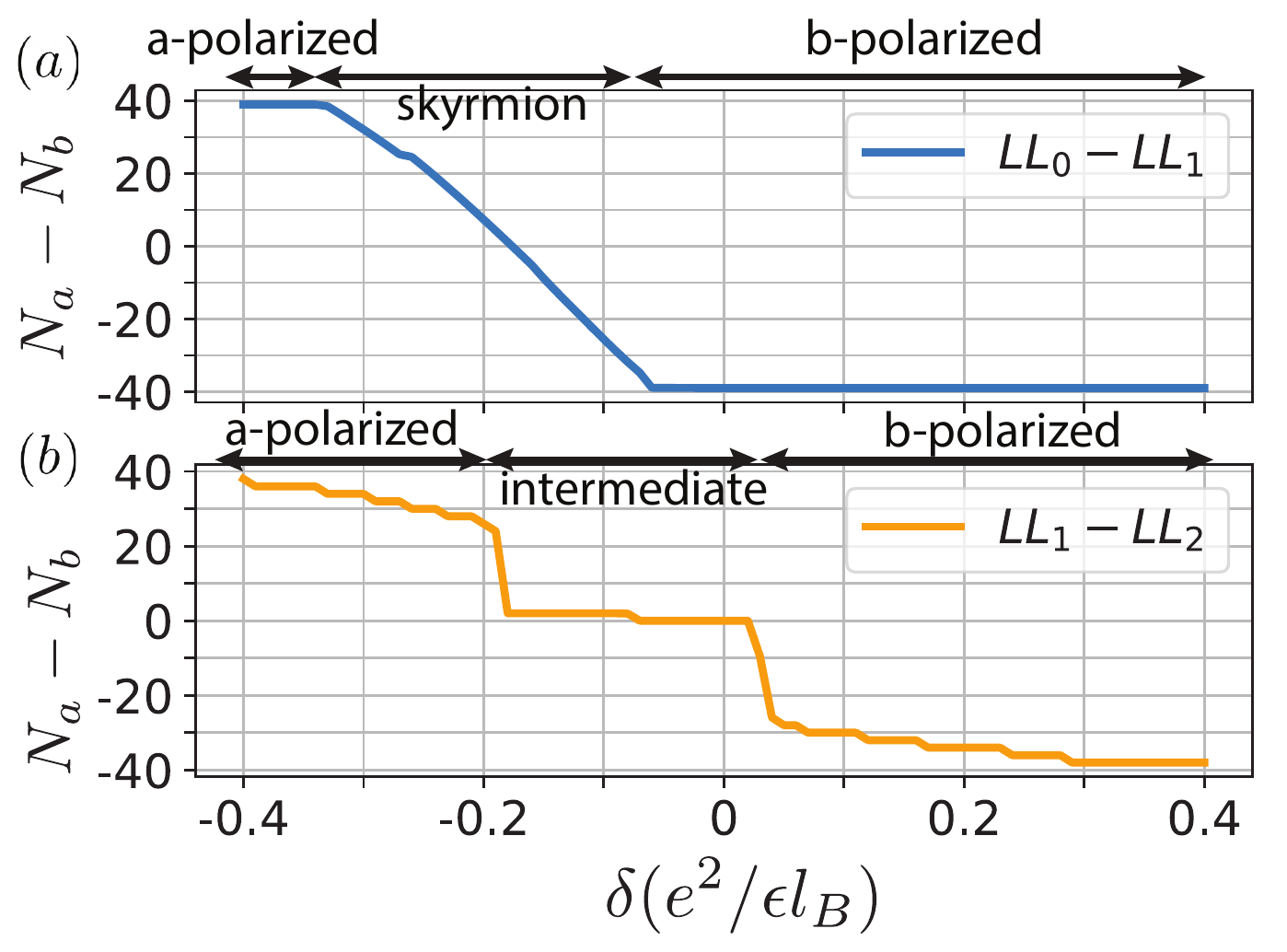}
  \caption{ {\bf Layer occupation.} The occupation difference $N_a-N_b = \sum_m \langle a^\dagger_m a_m \rangle - \langle b^\dagger_m b_m \rangle$ in a synthetic quantum Hall bilayer is plotted as a function of the detuning $\delta$.  (a) The bilayer system is obtained from coupling ($\Omega=10^{-4} e^2/\epsilon l_B$) between $LL_0$  and $LL_1$. (b) The bilayer system is obtained from coupling between $LL_1$ and $LL_2$. We consider $M=40$ states per Landau level, with $N=M+1$ electrons, and initialize the self-consistent iteration scheme with non-zero coherence $\langle a_{m}^\dagger b_{m+1} \rangle$ and $\langle b_{m+1}^\dagger a_{m}\rangle$. For sufficiently strong detuning, the system is trivially polarized in the manifold favored by the detuning. When $LL_0$ and $LL_1$ are coupled, the $b$-polarized phase extends to the regime of weak negative detuning due to the level-dependent interactions. Between the a-polarized and the b-polarized. The system exhibits the skyrmion phase for the $LL_0-LL_1$ whereas for the $LL_1-LL_2$ system, an intermediate minimally polarized phase is found. 
  \label{na_vs_det}
  }
\end{figure}
A first indication of skyrmionic behavior can be seen from the layer occupation $N_a-N_b = \sum_m (\langle a^\dagger_m a_m \rangle - \langle b^\dagger_m b_m \rangle)$ as a function of the detuning $\delta$, as illustrated in Fig. \ref{na_vs_det}. We consider both the $LL_0-LL_1$ synthetic bilayer, and the $LL_1-LL_2$ synthetic bilayer. Obviously, both systems exhibit highly polarized phases for sufficiently large $|\delta|$. Interestingly, when $LL_0$ and $LL_1$ are coupled, the $b$-polarized phase non-trivially extends into the regime of negative detuning. This already indicates that, in this regime, the layer polarization is not a single-particle effect, but due to the fact that the Coulombic repulsion is most efficiently minimized when the majority of particles occupy the $n=0$ Landau level (the $b$-level). On the other hand, for the case of $LL_1-LL_2$ coupling, such a phase with interaction-induced polarization is absent. 

Another difference between the $LL_0-LL_1$ bilayer and the $LL_1-LL_2$ bilayer can be seen from Fig. \ref{na_vs_det}: While for $LL_0-LL_1$ coupling, the occupation $N_a$ linearly increases as the detuning $\delta$ is decreased, the behavior in the $LL_1-LL_2$ bilayer exhibit abrupt jumps. At $\delta\approx 0$, the system jumps from an $b$-polarized phase into an (almost) unpolarized phase. At $\delta\approx -0.15 e^2/\epsilon l_B$, it jumps from this unpolarized phase into the $a$-polarized phase. This behavior is not consistent with a skyrmionic texture, which would allow for a continuous depolarization of the system. Hence, from the behavior of the layer occupation, we may already expect that skyrmions are supported by the $LL_0-LL_1$ bilayer, but not by the $LL_1-LL_2$ bilayer.

\paragraph{Orbital occupation.}
\begin{figure}[h]
 \includegraphics[width=1\columnwidth]{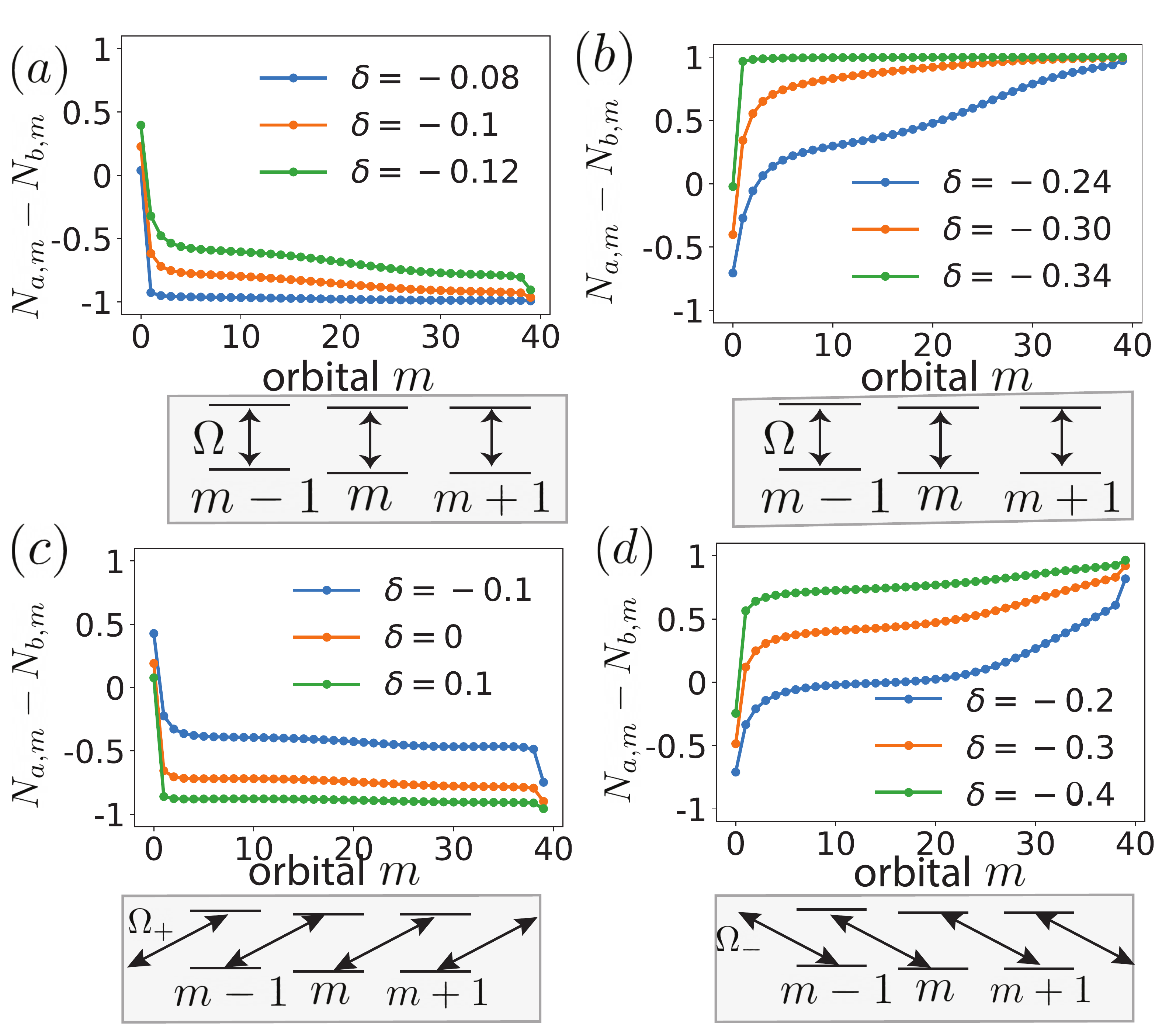}
  \caption{ {\bf Orbital occupation.}
  We plot the population difference $N_{a,m} - N_{b,m} = \langle a^\dagger_m a_m \rangle - \langle b^\dagger_m b_m \rangle$ between the two synthetic layers for each orbital $m$, for different values of the detuning $\delta$ (in units $e^2/\epsilon l_B$) The system is a $LL_0-LL_1$ synthetic bilayer, consisting of $M=40$ orbitals filled with $N=M+1$ electrons.
  In (a), the detuning favors polarization in the $b$-manifold, and accordingly,} the skyrmionic solution is triggered by dominant coherences $\langle a^\dagger_m b_{m-1}\rangle$ and $\langle b^\dagger_{m-1} a_{m}\rangle$ in the initial guess. In (b), the situation is opposite, as the detuning favors the $a$-manifold, and coherences $\langle a^\dagger_m b_{m+1}\rangle$ and $\langle b^\dagger_{m+1} a_{m}\rangle$ have to be chosen. In both panels, the Rabi coupling is $\Omega=10^{-4} e^2/\epsilon l_B$, and it connects orbitals with equal $m$, as schematically indicated below each of the plots. In panel (c) and (d), all initial coherences are chosen to be very weak ($\sim 10^{-5}$) and random, and the skyrmionic solution is now triggered through coupling of photons with orbital angular momentum $\ell=\pm \hbar$, with $\Omega_+ = 0.05e^2/\epsilon l_B $ in (c), and $\Omega_- = 0.05e^2/\epsilon l_B $ in (d).  As illustrated below the plots, such an optical coupling connects orbitals $m$ and $m\pm1$.
  In all four panels, the blue curves are chosen closer to the Zeeman-polarized regime, and the extra particle affects the polarization of only a few orbitals. In contrast, the green curves (which are furthest away from the Zeeman-polarized regime) show that most orbitals throughout the system become depolarized.
  \label{rhospin}
  
\end{figure}

In the Fig. \ref{rhospin}, we plot the orbital occupation difference $N_{a,m}-N_{b,m} =  \langle a^\dagger_m a_m \rangle - \langle b^\dagger_m b_m \rangle$, where $m$ is the orbital angular momentum. It reveals how the layer polarization changes locally, as one moves from the center (small $m$) to the edge (large $m$) of the system. When the detuning is chosen closer to a Zeeman-polarized regime (blue curves), only the orbital in the center becomes depolarized from the presence of an extra electron (on top of filling 1). The extra particle behaves like a single-particle excitation. In contrast, when the effect of Zeeman-polarization becomes weaker (green and orange curves), more orbitals become depolarized or even oppositely polarized through the presence of the extra particle. In these cases, the layer polarization winds from one polarization to the opposite polarization, as one moves through the system. The extra particle behaves like a skyrmion.

In Fig. \ref{rhospin}(a), close to the $b$-polarized regime, the skyrmion is obtained by choosing non-zero coherence $\langle a^\dagger_m b_{m-1}\rangle$ and $\langle b^\dagger_{m-1} a_{m}\rangle$ in the initial guess, and the system winds from an $a$-polarization in the center to $b$-polarization at the edge. For smaller values of $\delta$, when the system comes closer to the $a$-polarized paramagnetic phase, this kind of skyrmion becomes instable. Instead, we then obtain solutions with opposite winding behavior, as shown in panel (b). These solutions are obtained from non-zero coherences $\langle a^\dagger_m b_{m+1}\rangle$ and $\langle b^\dagger_{m+1} a_{m}\rangle$. 

In both cases, Fig. \ref{rhospin}(a) and (b), the Rabi frequency must be chosen sufficiently small ($\Omega \sim 10^{-4} e^2/\epsilon l_B$) in order to obtain skyrmionic solutions.  Strikingly, for the synthetic bilayer there is a relatively simple way of stabilizing the spin textures in the presence of stronger coupling: This can be achieved by replacing $\frac{\Omega}{2}(a_m^\dagger b_m + {\rm h.c.} )$ with  $\frac{\Omega_\pm}{2}(a_m^\dagger b_{m\pm1} + {\rm h.c.} )$, that is, by applying a coupling with photons with orbital angular momentum $\ell=\pm \hbar$. Such a strategy has already been suggested to create topological defects in chiral magnets \cite{fujita17}. As we show in panel (c) and (d) of Fig. \ref{rhospin}, the OAM coupling leads to very similar spin textures as in panels (a) and (b). Notably, the self-consistent equations now converge to this solution even without imposing them in the initial guess, and the spin textures remain present even for strong Rabi couplings, $\Omega_{\pm} \sim 0.05 e^2/\epsilon l_B$.

\paragraph{$LL_1-LL_2$ bilayer.}
The behavior of the $LL_1-LL_2$ bilayer is found to be quite different. In this case, skyrmionic correlations are fully suppressed. This is true both for the situation where we initialize the system in a state with non-zero skyrmionic correlations, and for the case where a coupling $\Omega_\pm$ is applied. The orbital populations will then remains close to zero throughout the system, with approximate the same population for all of the $2M$ orbitals. The dominant coherences established in the $LL_1-LL_2$ bilayer are of the form $\langle a_m^\dagger a_{m\pm1} \rangle$ and $\langle b_m^\dagger b_{m\pm1} \rangle$. These correlations indicate antiferromagnetic ordering: If an $a$-type ($b$-type) particle occupies an orbital $m$, the neighboring orbitals $m\pm1$ are unlikely to be populated by the same type of particle. Notably, these coherences acquire large non-zero values even if they are initially set to zero. This is possible only due to finite machine precision.

The absence of spin textures in the $LL_1-LL_2$ bilayer is not very surprising if one recalls the tendency of singlet formation due to its peculiar Haldane pseudopotentials. Such interactions prevent the formation of a quantum Hall ferromagnet at $\nu=1$, and thus, of skyrmionic excitations in the presence of $N=M+1$ electrons. This behavior illustrates, once more, that the $LL_0-LL_1$ bilayer and the $LL_1-LL_2$ bilayer behave in completely different ways, despite being seemingly very similar systems.

\subsection{Exact diagonalization}
\begin{figure}[h]
 \includegraphics[width=1\columnwidth]{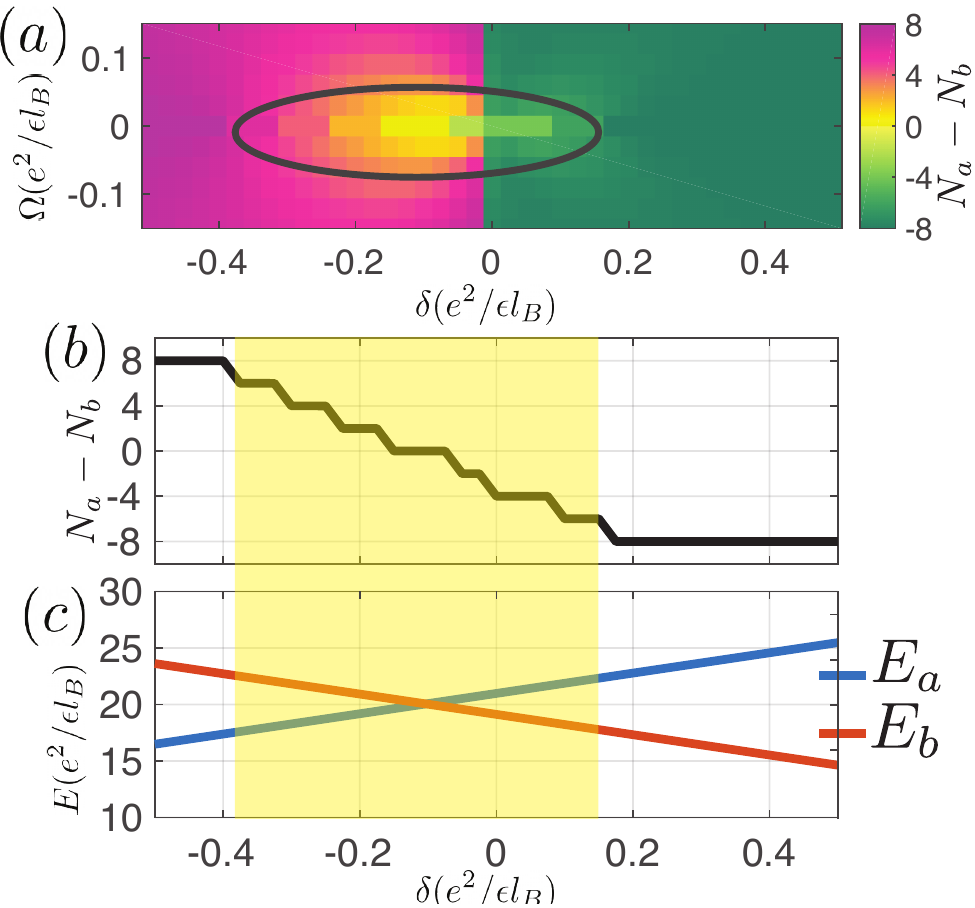}
  \caption{(a) The occupation difference $N_a-N_b$ as function of detunning $\delta$ and Rabi frequency $\Omega$. As $\delta \approx -0.1$ and $\Omega \approx 0$, the occupation difference $N_a-N_b$ reaches zero, which indicates the skyrmion phase. In the simulation, we choose  $N_e = 10$, $N_{\phi} = 9$.  The skyrmion phase is enclosed in the black circle. (b) The occupation difference $N_a-N_b$ as function of $\delta$ with $\Omega = 10^{-4}(e^2/\epsilon l_B^2)$. The skyrmion phase is highlighted in the yellow region. (c)  Coulomb interaction energy for the filled Landau level as function detunning $\delta$. At $\delta = 0$, the Coulomb interaction energy of b-level is intrinsically lower than the a-level.  The system exhibits the skyrmion phase when the detunning balance the unequal Coulomb interaction energy. In the simulation, we choose $N_e = 9$, $N_{\phi} = 9$.
\label{skyrmion_ED}}
\end{figure}

In order to back our mean-field calculation, we have also performed  exact diagonalization on the spherical geometry.  We assume that the number of electrons is $N_e = 10$ and the total number of quantum fluxes $N_\phi = 9$. Here, we consider $LL_0-LL_1$ coupling. In the large Zeeman energy limit, the system energetically favors a state with a single spin flip. As the Zeeman energy decreases, the system undergoes a phase transition to the skyrmion phase. To explore the phase diagram in the regime where the detuning $\delta$ and the Rabi frequency $\Omega$ are comparable, we calculate the number difference between the dressed level $N_a - N_b = \langle\tilde{\tau}^z_{0,m}\rangle$ as shown in Fig. \ref{skyrmion_ED}(a). 

In the conventional bilayer quantum Hall system whith $SU(2)$ symmetric interactions, the system possesses large skyrmion excitations when the Zeeman energy vanishes. However, in the synthetic quantum Hall bilayer system, the largest skyrmion excitation occurs with the finite negative detunning, at around $\delta=-0.1$. This is in agreement with the behavior found in the mean-field calculation (cf. Fig. \ref{na_vs_det}), indicating that this behavior is independent from the size of the system.

To understand the interplay between Coulomb interaction and the Zeeman energy in the synthetic bilayer graphene system, we show the ground state energy of the filled zeroth(first) Landau level as function of detunning in Fig. \ref{skyrmion_ED}(c). When the detunning $\delta = 0$, we observe that the ground state energy of filled zeroth Landau level is higher than the filled first Landau level due to the Coulomb interaction. The unbalanced Coulomb energy competes with the formation of skyrmion. By decreasing the detunning, the energy of the two filled Landau level becomes the same.  When the two Landau levels are energetically equally favorable, the size of the skyrmion reaches maximum.

\section{Summary and Outlook}
In summary, we have demonstrated that the the laser field coupled to the single-layer graphene provides a versatile platform to study the bilayer quantum Hall physics. By using the infinite density matrix renormalization group and exact diagonalization, we show that the system exhibits the bilayer Fibonacci phase which can be of interest for topological quantum computation with its non-abelian anyonic statistic. Moreover, we also explore the phase diagram for topological spin texture excitations known as skyrmion phase in the quantum Hall ferromagnetic regime. Apart from providing a synthetic bilayer structure, optical coupling between Landau levels may also enable the controlled engineering of three-body interaction terms from second-order transition processes. Future work on optically driven quantum Hall systems may explore this interesting scenario.

\acknowledgments{ 
TG acknowledges financial support from a fellowship granted by “la Caixa” Foundation (ID 100010434, fellowship code LCF/BQ/PI19/11690013), as well as funding
from the Spanish Ministry MINECO (National Plan 15 Grant: FISICATEAMO No. FIS2016-79508-P, SEVERO OCHOA No. SEV-2015-0522, FPI), European Social Fund, Fundaci\'o Cellex, Fundaci\'o Mir-Puig, Generalitat de Catalunya (AGAUR Grant No. 2017 SGR 1341,CERCA/Program), ERC AdG NOQIA, EU FEDER, MINECO-EU QUANTERA MAQS (funded by The State Research Agency (AEI) PCI2019-111828-2 / 10.13039/501100011033), and the National Science Centre, Poland-Symfonia Grant No. 2016/20/W/ST4/00314. ZP and MH were supported by ARO, AFOSR-MURI and Physics Frontier Center at Joint Quantum Institute. Z.L. is supported by the National Natural Science Foundation of China through Grant No. 11974014.

\bibliography{ref.bib}

\end{document}